\documentstyle[12pt]{article}
\title{A Phenomenological Analysis of Higher Fock State
Contributions to the $\chi_{cJ}$ Decays\thanks{
 This work is in part supported by the National Science Foundation of China.}}

\vspace{3cm}

\author{Tao Huang and Huifang Wu   \\
        Institute of High Energy Physics, P.O.Box 918 ,Beijing 100039, P. R. China\\
         and \\
        Institute of Theoretical Physics, P.O.Box 8730,Beijing 100080, P. R.China}
\date{}
\begin{document}
\maketitle

\begin{abstract}
We present a phenomenological analysis of higher
Fack state contributions to the $\chi_{cJ}$ decays by using the
recent BES experimental data. It is found that the higher Fock
state $\mid (c\bar{c})_{8}g>$ makes an important contributions to
the inclusive and exclusive processes with respect to that from
the valence Fock state $\mid c\bar{c} >$ of the $\chi_{cJ}$ and
some constraints of these contributions are obtained for the
$\chi_{c0}$ and $\chi_{c2}$ states in order to fit the
experimental data.\\
\vspace{2cm}

\vspace{0.5cm}

\end{abstract}

\newpage
\section*{1. Introduction}

{\hskip 0.6cm}Recently, BES has reported about 30 channels of $\chi_{cJ}$
hadronic decay$^{[1]}$ and the decays to $\pi \pi, KK$ among
them have improved greatly with high precision$^{[2]}$. For
example, some branching ratios are shown in the following table.

\vspace{1cm}

\begin{center}
\begin{tabular}{|c|c|} \hline
Decay channels &BES $(\times 10^{-3})$ \\ \hline
$~~~~~~~~Br(\chi_{c0} \rightarrow \pi^{+}\pi^{-})~~~~~~~~~~~$
&~~~~~~~~$4.68 \pm 0.26 \pm0.65~~~~~~~~~$\\ \hline
$Br(\chi_{c2} \rightarrow \pi^{+}\pi^{-})$ &$1.49 \pm 0.14 \pm
0.22$\\ \hline
$Br(\chi_{c0} \rightarrow K^{+}K^{-})$ &$5.68 \pm 0.35 \pm
0.85$\\ \hline
$Br(\chi_{c2} \rightarrow K^{+} K^{-})$ &$0.79 \pm 0.14 \pm
0.13$\\ \hline
\end{tabular}
\end{center}

\vspace{1cm}

$\chi_{cJ}$ are the P-wave bound state of $c\bar{c}$ and can be
treated approximately as a non-relativistic system due to the
large mass of the charm quark. In such a system, the charm quark
and anti-charm quark are coupled to light quarks by two-gluon
exchange which has large momentum flow. Then these light quarks
will build up the products of the decay, i.e. light hadrons. The
total hadronic widths of $\chi_{cJ}$ are approximately equal to
the widths of the decays into two gluons$^{[3]}$,
\begin{eqnarray}
\Gamma^{(1)}_{tot}(\chi_{c0}) =
\frac{6~\alpha^{2}_{s}(M^{2}_{c})}{M^{4}_{c}} \mid R{'}_{p}(0)
\mid^{2}
\end{eqnarray}

\begin{eqnarray}
\Gamma^{(1)}_{tot}(\chi_{c2}) =
\frac{8~\alpha^{2}_{s}(M^{2}_{c})}{5M^{4}_{c}} \mid R^{'}_{p}(0)
\mid^{2}
\end{eqnarray}
where $R^{'}_{p}(0)$ is the derivative of the radial wavefunction
of the heavy quankonium at origin. Here we only consider the
valance Fock state contribution. Theorefore the ratio of the total
decay widths
\begin{eqnarray}
R \equiv \frac{\Gamma_{tot}(\chi_{c0})}{\Gamma_{tot}(\chi_{c2})}\cong
\frac{\Gamma^{(1)}_{tot}(\chi_{c0})}{\Gamma^{(1)}_{tot}(\chi_{c2})}=\frac{15}{4}
\end{eqnarray}
after neglecting the $v^{2}$ corrections. It was assumed
that the single non-perturbative quantity $R^{'}_{p}(0)$ includes
the long-distance effects. However, Barbieri et al.$^{[4]}$ found
that the coefficients of $\mid R^{'}_{p}(0) \mid^{2}$ depend
logarithmically on an infrared cutoff on the energies of the
final-state gluons. Bodwin et. al.$^{[5]}$ pointed out that to
calculate the coefficients of $\mid R^{'}_{p}(0)\mid^{2}$ to
relative order $\alpha^{n}_{s}(\alpha_{s}(Mv))$ should include all
operators whose matrix elements are of relative order $v^{n}$ or
less. Thus the annihilation rate of the $\chi_{cJ}$ into light
hadrons becomes$^{[6]}$
\begin{eqnarray}
\Gamma(\chi_{cJ}) = \sum_{n} \frac{2 ~Im ~fn(\Lambda)}{M^{d_n -
4}} <  \chi_{cJ} \mid O_{n} (\Lambda) \mid \chi_{cJ} >
\end{eqnarray}
and the dependence on the arbitrary factorization scale $\Lambda$
in Eq.(4) cancels between the coefficients and the operators. The
P-wave $\chi_{cJ}$ state can be expressed as
\begin{eqnarray}
\mid \chi_{cJ} > = O(1) \mid (c\bar{c})_{1} (^{3}P_{J}) >
+ O(v) \mid (c\bar{c})_{8} (^{3}S_{1}) g > + ~~
O(v^{2})
\end{eqnarray}
As keeping the first term in Eq.(5), Eq.(4) becomes Eqs.(1)
and (2). They also show that the inclusion of the higher Fock
states in the factorization formulas removes the dependence of the
decay rate on an arbitrary infrared cutoff. Eq.(4) gives the
corrections to the leading contribution Eqs.(1) and (2).

Now we consider the case of exclusive decays of $\chi_{cJ}$, such
as $\chi_{cJ} \rightarrow \pi \pi, K K$,... . Following the
framework of calculations of exclusive processes at large momentum
transfers$^{[7]}$, the decay amplitude of $\chi_{cJ}$ can be
factorized into two parts: a hard amplitude $T_{H}$ calculable in
perturbative QCD, and the distribute amplitudes $\phi_{H}$ for
each hadron H. For example, taking the valence Fock state in
Eq.(5), the decay width of $\chi_{cJ} \rightarrow \pi\pi$ is given
by
\begin{eqnarray}
\Gamma^{(1)}(\chi_{c0} \rightarrow \pi\pi) = \frac{C^{2}[4\pi
~\alpha_{s}(M^{2})]^{4} }{2\pi^{2}\cdot 32M^{8}} \mid
R^{'}_{p}(0)\mid^{2} ~~\mid I^{\pi}_{0} \mid^{2}
\end{eqnarray}
and
\begin{eqnarray}
 \Gamma^{(1)}(\chi_{c2} \rightarrow \pi\pi) = \frac{C^{2}[4\pi
~\alpha_{s}(M^{2})]^{4} }{5\pi^{2}\cdot 32M^{8}} \mid
 R^{'}_{p}(0)\mid^{2} ~~\mid I^{\pi}_{2} \mid^{2}
\end{eqnarray}
where
\begin{eqnarray}
I^{\pi}_{0} = \int^{1}_{0} dx \int^{1}_{0} dy \phi_{\pi}(x,Q^{2})
\frac{2+\frac{(x-y)^{2}}{x+y-2xy}}{x(1-x)(x+y-2xy)y(1-y)}
\phi_{\pi}(y,Q^{2})
\end{eqnarray}

\begin{eqnarray}
I^{\pi}_{2} = \int^{1}_{0} dx \int^{1}_{0} dy \phi_{\pi}(x,Q^{2})
\frac{1-\frac{(x-y)^{2}}{x+y-2xy}}{x(1-x)(x+y-2xy)y(1-y)}
\phi_{\pi}(y,Q^{2})
\end{eqnarray}
and
\begin{eqnarray}
C = (\frac{1}{\sqrt{n_c}})^{3} \sum_{AB} [Tr(T^{A} T^{B})]^{2} =
\frac{2}{3\sqrt{3}}
\end{eqnarray}
is the color factor. The distribution amplitude
$\phi_{\pi}(x,Q^{2})$ in Eqs.(8) and (9) is defined by the valence
wave function $\psi_{q\bar{q} / \pi} (x,k_{\bot})$ in the
light-cone$^{[8]}$,

\begin{eqnarray}
\phi_{\pi}(x, Q^{2}) = \int^{Q^{2}}_{0} d^{2} k_{\bot}
~\psi_{q\bar{q}/\pi} (x, k_{\bot}) ~~.
\end{eqnarray}

Usually, one assumes that higher Fock state contributions are
suppressed by v with respect to that from the valence Fock state
$\mid (c\bar{c})_{1} >$. It was shown$^{[6]}$ this suppression for
inclusive decays of $\chi_{cJ}$ does not hold and the contrbution
from the color-octet $\mid (c\bar{c})_{8} (^{3}S_{1}) g >$ (i.e.
the one arising from the higher Fock state $\mid c\bar{c}g >$ with
$c\bar{c}$ in a lolor-octet state) in Eq.(5) is not suppressed by
v at all. Fothermore, Ref.[9] argue that for exclusive $\chi_{cJ}$
decays the color-octet contribution is not suppressed by powers of either
v or $\frac{1}{M_{c}}$ too.

After taking into account for the higher Fock state contribution we can
write down the decay width of the process $\chi_{cJ} \rightarrow
\pi\pi$,
\begin{eqnarray}
\Gamma(\chi_{cJ} \rightarrow \pi\pi) = \Gamma^{(1)} (\chi_{cJ}
\rightarrow \pi\pi) + \Gamma^{(8)} (\chi_{cJ} \rightarrow \pi\pi)
~~,
\end{eqnarray}
where $\Gamma^{(8)}(\chi_{cJ} \rightarrow \pi\pi)$ is determined
by the color-octet contribution. It will be estimated by the wave
function of the color-octet $\mid (c\bar{c})_{8} g >$ and there
are more uncertainties to calculate the color-octet contribution.

\section*{2. Phenomenological analysis to the decay widths}

{\hskip 0.6cm}It is shown that the color octet of $\chi_{cJ}$ will
make contributions to the decay width of them. However we haven't
a good framework to calculate their contributions precisely yet at
present. In order to estimate them phenomenologically we rewrite
the decay widths of $\chi_{cJ}$ as
\begin{eqnarray}
\Gamma(\chi_{cJ}) = \Gamma^{(1)} (\chi_{cJ}) [1+\Delta_{J}]
\end{eqnarray}

\begin{eqnarray}
\Gamma(\chi_{cJ} \rightarrow \pi\pi) = \Gamma^{(1)} (\chi_{cJ}\rightarrow \pi\pi)
[1+ a^{\pi}_{J}]
\end{eqnarray}

\begin{eqnarray}
\Gamma(\chi_{cJ} \rightarrow KK) = \Gamma^{(1)} (\chi_{cJ}\rightarrow KK)
[1+ a^{K}_{J}]
\end{eqnarray}
where $\Gamma^{(1)}(\chi_{cJ}), \Gamma^{(1)}(\chi_{cJ} \rightarrow
\pi\pi)$ and $\Gamma^{(1)}(\chi_{cJ} \rightarrow KK)$
are determined by the color-singlet state of $\chi_{cJ}$. All of
contributions from the color-octet state of $\chi_{cJ}$ are
included in the parameters $\Delta_{J}, a^{\pi}_{J}$ and
$a^{K}_{J}$.  $\Gamma^{(1)}(\chi_{c0}$ and
$\Gamma^{(1)}(\chi_{c2}$ are determined by Eqs.(1) and (2). Then
we have the raio R
\begin{eqnarray}
R \equiv \frac{\Gamma_{tot}(\chi_{CO})}{\Gamma_{tot}(\chi_{c2})}
&=&
\frac {\Gamma^{(1)}_{tot}(\chi_{CO})(1+\Delta_{0})}{\Gamma^{(1)}_{tot}(\chi_{c2})(1+\Delta_{2})}
 \nonumber \\
&=& \frac{15}{4} ~~\frac{(1+\Delta_{0})}{(1+\Delta_{2})}
\end{eqnarray}
and the branching ratios
\begin{eqnarray}
Br(\chi_{c0} \rightarrow \pi^{+} \pi^{-}) = \frac{8 \pi^{2}}{81} ~~
\frac{\alpha^{2}_{s}(M^{2}_{c})}{M^{4}_{c}} \mid I^{\pi}_{0} \mid^{2}
\frac{1+a^{\pi}_{0}}{1+\Delta_{0}}
\end{eqnarray}

\begin{eqnarray}
Br(\chi_{c2} \rightarrow \pi^{+} \pi^{-}) = \frac{4 \pi^{2}}{27} ~~
\frac{\alpha^{2}_{s}(M^{2}_{c})}{M^{4}_{c}} \mid I^{\pi}_{2} \mid^{2}
\frac{1+a^{\pi}_{2}}{1+\Delta_{2}}
\end{eqnarray}

\begin{eqnarray}
Br(\chi_{c0} \rightarrow K^{+} K^{-}) = \frac{8 \pi^{2}}{81} ~~
\frac{\alpha^{2}_{s}(M^{2}_{c})}{M^{4}_{c}} \mid I^{K}_{0} \mid^{2}
\frac{1+a^{K}_{0}}{1+\Delta_{0}}
\end{eqnarray}

\begin{eqnarray}
Br(\chi_{c2} \rightarrow K^{+} K^{-}) = \frac{4 \pi^{2}}{27} ~~
\frac{\alpha^{2}_{s}(M^{2}_{c})}{M^{4}_{c}} \mid I^{K}_{2}
\mid^{2}
\frac{1+a^{K}_{2}}{1+\Delta_{2}} ~~.
\end{eqnarray}
Thus we eliminate the uncertainty of $R^{'}_{p}(0)$ in
Eqs.(16-20). In fact, the parameters $\Delta_{J}, a^{\pi}_{J}$ and $a^{K}_{J}$ may
include all of contributions from the higher Fock state and the
higher order terms. From the experiments
\begin{eqnarray*}
\Gamma_{tot}(\chi_{c0}) &=& 14.3 \pm 2.0 \pm 3.0 ~MeV  \\
\Gamma_{tot}(\chi_{c2}) &=& 2.00 \pm 0.18 ~MeV
\end{eqnarray*}
follows
\begin{eqnarray}
\frac{1+\Delta_{0}}{1+\Delta_{2}}  \sim 1 - 3
\end{eqnarray}
which leads an inequality
\begin{eqnarray}
\Delta_{0} > \Delta_{2}
\end{eqnarray}

Recent BES measurement gets the ratios of the branching fractions,
\begin{eqnarray*}
\frac{Br(\chi_{c0} \rightarrow \pi^{+}\pi^{-})}{Br(\chi_{c0}
\rightarrow K^{+} K^{-})} = 0.82 \pm 0.15
\end{eqnarray*}
and
\begin{eqnarray*}
\frac{Br(\chi_{c2} \rightarrow \pi^{+}\pi^{-})}{Br(\chi_{c2}
\rightarrow K^{+} K^{-})} = 1.88 \pm 0.51
\end{eqnarray*}
Comparing the experimental data with Eqs.(17-20), we can obtain
some constrants on the parameters $a^{\pi}_{J}$ and
$a^{K}_{J}$,
\begin{eqnarray}
\frac{\mid I^{\pi}_{0} \mid^{2} (1+a^{\pi}_{0})}{\mid I^{K}_{0} \mid^{2}
(1+a^{K}_{0})} &=& 0.82 \pm 0.15
\end{eqnarray}

\begin{eqnarray}
\frac{\mid I^{\pi}_{2} \mid^{2} (1+a^{\pi}_{2})}{\mid I^{K}_{2} \mid^{2}
(1+a^{K}_{2})} = 1.88 \pm 0.51
\end{eqnarray}
If we ignore the SU(3) symmetry breaking of the $\pi$ and $K$
wave functions, Eqs.(23) and (24) tell us one constraint
\begin{eqnarray}
a^{\pi}_{0} \simeq a^{K}_{0}
\end{eqnarray}
for the $\chi_{c0}$ state and another constraint
\begin{eqnarray}
a^{\pi}_{2} > a^{K}_{2}
\end{eqnarray}
for the $\chi_{c2}$ state. Therefore the present experimental data
have put some constraint on the color-octet contributions. In
order to fit the BES data it requires that (1) the
contributions to the $\chi_{c0}$ inclusive decay are larger than
the contributions to $\chi_{2}$ from the higher Fock states i.e.
$\Delta_{0} > \Delta_{2}$; (2) For the $\chi_{c0}$ state, the
corrections to the process $\chi_{c0} \rightarrow \pi \pi$ from the
higher Fock states is approximately equal to the correcitons to
the process $\chi_{c0} \rightarrow KK$; (3) For the
$\chi_{c2}$ states, the corrections to the process $\chi_{c2}
\rightarrow \pi\pi$ is larger than the corrections to the process
$\chi_{2} \rightarrow KK$ from the higher Fock states.

\section*{3. Numerical estimate to the valence and hgiher Fock
state contribution}

{\hskip 0.6cm}Now let us calculate the valence Fock state
contribution, such as $\Gamma^{(1)}(\chi_{cJ})$, $\Gamma^{(1)}
(\chi_{cJ} \rightarrow \pi\pi)$ and $\Gamma^{(1)} (\chi_{cJ}
\rightarrow KK)$ in the leading order. The results,
according to the Eqs.(7-9), are all dependent on the distribution
amplitude $\phi(x,Q^{2})$ of the pion and kaon which are
independent of the processes apart from the energy scale $Q^{2}$.
the $\phi_{M}(x, Q^{2})$ of the meson is defined by Eq.(11) and
$\psi_{q\bar{q} /M } (x, k_{\bot} )$ is the valance wavefunction
of the meson in the light-cone framework. The general solution of
the QCD evolution equation$^{[8]}$
\begin{eqnarray}
\phi_{M}(x, Q^{2}) = x(1-x) \sum^{\infty}_{n=0} a_{n}(Q^{2}_{0})
C^{3/2}_{n} (2x
-1)(\frac{\alpha_{s}(Q^{2})}{\alpha_{s}(Q^{2}_{0})} )^{\gamma_{n}}
\end{eqnarray}
and the coefficients $a_{n}(Q^{2}_{0})$ are dependent upon the
initial distribution amplitude $\phi_{M}(x, Q^{2}_{0})$ through its
definition
\begin{eqnarray}
a_{n}(Q^{2}_{0}) = \frac{2(2n + 3)}{(2+n)(1+n)} \int^{1}_{-1}
d(2x-1)C^{3/2}_{n}(2x-1)\phi_{\pi}(x, Q^{2}_{0})
\end{eqnarray}
with $a_{0} = \sqrt{3} f_{M}$. $f_{M}$ is the decay constant of
the meson. As $Q^{2} \rightarrow \infty, \phi_{M}(x, Q^{2})$ goes
to the asymptotic form $\phi^{as}_{M}(x)$
\begin{eqnarray}
\phi^{as}_{M}(x) = \sqrt{3} ~f_{M} x(1-x) ~~.
\end{eqnarray}
If employing the asymptotic form, one finds the asymptotic values
of the branching ratios of the $\chi_{cJ}$ are much smaller than
the experimental values. In order to fit the data one
possible solution is to take a much
wider distribution amplitude$^{[10]}$. For examle, from the
Chernyak-Zhitnisky (CZ) form$^{[11]}$
\begin{eqnarray}
\phi_{M}(x) = \frac{f_{M}}{\sqrt{3}} 15(2x-1)^{2} x(1-x)
\end{eqnarray}
the valence Fock state contribution can be consistant
with experimental values very well without any
color-octet contribution. However the recent studies on the
processes involving the pion disfavor the CZ distribution
amplitude$^{[12-15]}$. It follows that the pion distribution
amplitude is close to the asymptotic form. Following Refs.[12,16],
we take the model wavefunction
\begin{eqnarray}
\psi_{\pi}(x,k_{\bot}) = A_{\pi} ~~exp[-\frac{k^{2}_{\bot} + m^{2}}{8
\beta^{2} x(1-x)}]
\end{eqnarray}
and
\begin{eqnarray}
\psi_{K}(x, k_{\bot}) = A_{K} ~~exp [- (\frac{k^{2}_{\bot} + m^{2}}{8
 \beta^{2} x} + \frac{k^{2}_{\bot} + m^{2}_s}{8 \beta^{2}(1-x)})]
\end{eqnarray}
with $A_{\pi} = 32 GeV^{-1}, A_{K} = 52.6 GeV^{-2}, \beta = 385 MeV, m = 298 MeV$ and
$m_{s} = 550 MeV$. It leads the distribution amplitudes
\begin{eqnarray}
\phi_{\pi}(x) = \frac{2\beta^{2} A_{\pi}}{(2\pi)^{2}} x(1-x) ~exp
(- \frac{m^{2}}{8 \beta^{2} x(1-x)} )
\end{eqnarray}
and
\begin{eqnarray}
\phi_{K}(x) = \frac{2\beta^{2} A_{K}}{(2\pi)^{2}} x(1-x) ~exp
(- \frac{1}{8\beta^{2}} (\frac{m^{2}}{x} + \frac{m^{2}_{s}}{1-x}))
\end{eqnarray}
Substituting Eqs.(33-34) into (7-9, 17-20) we can get the
following numerical results
\begin{eqnarray}
Br(\chi_{c0} \rightarrow \pi^{+} \pi^{-} ) &=& B_{0}(0.1109)^{2}
\frac{1+a^{\pi}_{0}}{1+\Delta_{0}}  \\
Br(\chi_{c2} \rightarrow \pi^{+}\pi^{-} ) &=& B_{2}(0.4538 \times 10^{-1})^{2}
\frac{1+a^{\pi}_{2}}{1+\Delta_{2}} \\
Br(\chi_{c0} \rightarrow K^{+} K^{-} ) &=& B_{0}(0.1256)^{2}
\frac{1+a^{K}_{0}}{1+\Delta_{0}} \\
Br(\chi_{c2} \rightarrow K^{+} K^{-} ) &=& B_{2}(0.5329 \times 10^{-1})^{2}
\frac{1+a^{K}_{2}}{1+\Delta_{2}}
\end{eqnarray}
where
\begin{eqnarray}
B_{0} = \frac{8\pi^{2}}{81}
~~\frac{\alpha^{2}_{s}(M^{2}_{c})}{M^{4}_{c}(GeV)^{-4}}, ~~
B_{2} = \frac{4\pi^{2}}{27}
~~\frac{\alpha^{2}_{s}(M^{2}_{c})}{M^{4}_{c}(GeV)^{-4}}
\end{eqnarray}
From Eqs.(35-38) follows
\begin{eqnarray}
\frac{Br(\chi_{c0} \rightarrow \pi^{+}\pi^{-})}
{Br(\chi_{c0} \rightarrow K^{+} K^{-})}
 = 0.78 \frac{1+a^{\pi}_{0}}{1+ a^{K}_{0}}
\end{eqnarray}
and
\begin{eqnarray}
\frac{Br(\chi_{2} \rightarrow \pi^{+}\pi^{-})}
{Br(\chi_{2} \rightarrow K^{+} K^{-})}
 = 0.73 \frac{1+a^{\pi}_{2}}{1+ a^{K}_{2}}
\end{eqnarray}
These ratios are independent of the parameters $\alpha_{s}(M_{c})$
and $M_{c}$. Comparing Eqs.(40-41) with BES experimental values
one obtains
\begin{eqnarray}
a^{\pi}_{0} \simeq a^{K}_{0}
\end{eqnarray}

\begin{eqnarray}
a^{\pi}_{2} \cong 2.58 ~~a^{K}_{2} + 1.58
\end{eqnarray}
which means that the higher Fock state contribution depends on the
total spin J of the p-wave state $\chi_{cJ}$. Eqs.(42-43)
show that $a^{\pi}_{2}$ is rather larger than $a^{K}_{2}$.

The precise prediction will be determined by the parameters
$\alpha_{s}(M_{c}), R^{'}_{p}(0), M_{c}, \Delta_{J}, a^{\pi}_{J}$
and $a^{K}_{J}$. In this paper we only apply the phenomenological
analysis to the branching ratios to eliminate the parameters
$\alpha_{s}(M_{c}), R^{'}_{p}(0)$ and $M_{c}$ for getting some
constraints on the parameters $\Delta_{J}, a^{\pi}_{J}$ and
$a^{K}_{J}$ .

\section*{4. Conclusion}

{\hskip 0.6cm}We have presented a phenomenological analysis of
higher Fock state contributions to the $\chi_{cJ}$ decays by using
the recent BES experimental data. In this paper we include the
higher Fock state contributions to the decays of $\chi_{cJ}$
beyond the valence Fock state contribution. We parameterize the
higher Fock state and higher order contributions as $\Delta_{J}$
for inclfusive processes and $a^{\pi}_{J}$ and $a^{K}_{J}$
for exclusive processes of the $\chi_{cJ}$. The recent
experimental data requires that (i) For the inclusive process of
the $\chi_{cJ}$, the higher Fock state $\mid (c\bar{c})_{8} g >$
makes an important contribution with respect to that from the
valance Fock state $\mid c\bar{c} >$ and they are different for
the $\chi_{c0}$ and $\chi_{c2}$ state, i.e. $\Delta_{0} >
\Delta_{2}$. (ii) The similar results are obtained for the
exclusive processes and we find $a^{\pi}_{0} \simeq
a^{K}_{0}$ for the $\chi_{c0}$ state and $a^{\pi}_{2} >
a^{K}_{2}$ for the $\chi_{c2}$ state.


\newpage
\begin{thebibliography}{99}
\bibitem{1} BES Collaboration, Nucl. Phys. 75B(1999)181; Z. P.
Zheng, Int. J. of Modem Phys. A15(2000)4723.
\bibitem{2} J. Z. Bai et al., Phys. Rev. Letts, 81(1998)3091.
\bibitem{3} V. A. Novirov et al., Phys. Rep. 41C(1978)1.
\bibitem{4} R. Barbieri, R. Gatto and E. Remiddi, Phys. Lett.
95B(1980)93; Nucl. Phys. B192(1981)61.
\bibitem{5} G. T. Bodwin, E. Braaten and G. P. Lepage, Phys. Rev.
D46(1992)R1914.
\bibitem{6} G. T. Bodwin, E. Braaten and G. P. Lepage, Phys. Rev.
D51(1995)1125.
\bibitem{7} A. Duncan and A. H. Mueller, Phys. lett. 93B(1980)119;
A. H. Mueller, Phys. Rep. 73C(1981)237.
\bibitem{8} G. P. Lepage and S. J. Brodsky, Phys. Rev.
D22(1980)2157.
\bibitem{9} J. Bolz, P. Kroll and G. A. Schuler, Phys. Letts
B392(1997)198; Eur. Phys. J. C2(1998)705.
\bibitem{10} X. N. Wang, X. D. Xiang and T. Huang, Commun. in
Theor. Phys. 5(1986)123.
\bibitem{11} V. L. Chernyak and A. R. Zhitnisky, Nucl. Phys.
B201(1982)492.
\bibitem{12} T. Huang, B. Q. Ma and Q. X. Shen, Phys. Phys. Rev.
D49(1994)1490 .
\bibitem{13} A. V. Radyushkin and R. T. Ruskov, Phys. Lett.
B374(1996)848.
\bibitem{14} R. Jakob, P. Kroll and M. Raulfs, J. Phys.
G22(1996)45; P. Kroll and M. Raulfs, Phys. Lett. B387(1996)848.
\bibitem{15} I. V. Musatov and A. V. Radyushkin, Phys. Rev.
D56(1997)2713.
\bibitem{16} S. J. Brodsky, T. Huang and G. P. Lepage, Particles
and Fields 2, eds Z. Capri and A. N. Kamal,(1982) p.143.
\end{thebibliography}
\end{document}